\documentclass[preprint,12pt]{elsarticle}

\usepackage{amssymb}
\usepackage{amsmath}
\usepackage{amsmath,amssymb,amsfonts}

\usepackage{graphicx}
\usepackage{textcomp}
\usepackage{algorithm}

\usepackage{algpseudocode}

\usepackage{algcompatible}
\usepackage{amsfonts}
\usepackage{booktabs}
\usepackage{setspace}
\journal{Nuclear Physics B}

\begin{document}

\begin{frontmatter}

%% Title, authors and addresses

%% use the tnoteref command within \title for footnotes;
%% use the tnotetext command for theassociated footnote;
%% use the fnref command within \author or \affiliation for footnotes;
%% use the fntext command for theassociated footnote;
%% use the corref command within \author for corresponding author footnotes;
%% use the cortext command for theassociated footnote;
%% use the ead command for the email address,
%% and the form \ead[url] for the home page:
%% \title{Title\tnoteref{label1}}
%% \tnotetext[label1]{}
%% \author{Name\corref{cor1}\fnref{label2}}
%% \ead{email address}
%% \ead[url]{home page}
%% \fntext[label2]{}
%% \cortext[cor1]{}
%% \affiliation{organization={},
%%             addressline={},
%%             city={},
%%             postcode={},
%%             state={},
%%             country={}}
%% \fntext[label3]{}

\title{GPDM: Generation-Prior Diffusion Model for Accelerated Direct Attenuation and Scatter Correction of Whole-body \textsuperscript{18}F-FDG PET}

%% use optional labels to link authors explicitly to addresses:
%% \author[label1,label2]{}
%% \affiliation[label1]{organization={},
%%             addressline={},
%%             city={},
%%             postcode={},
%%             state={},
%%             country={}}
%%
%% \affiliation[label2]{organization={},
%%             addressline={},
%%             city={},
%%             postcode={},
%%             state={},
%%             country={}}

\author[label1,label2,label3]{Min Jeong Cho} %% Author name
\author[label1,label2,label3]{Hyeong Seok Shim} %% Author name
\author[label3]{Sungyu Kim}
\author[label1,label2,label3,label4]{Jae Sung Lee}

\affiliation[label1]{organization={Interdisciplinary Program in Bioengineering},%Department and Organization
            addressline={Seoul National Graduate School College of Engineering}, 
            city={Seoul},
            postcode={08826}, 
            state={},
            country={South Korea}}

\affiliation[label2]{organization={Integrated Major in Innovative Medical Science},%Department and Organization
            addressline={Seoul National University Graduate School}, 
            city={Seoul},
            postcode={03080}, 
            state={},
            country={South Korea}}

\affiliation[label3]{organization={Department of Nuclear Medicine},%Department and Organization
            addressline={Seoul National University College of Medicine}, 
            city={Seoul},
            postcode={03080}, 
            state={},
            country={South Korea}}

\affiliation[label4]{organization={Brightonix Imaging Inc.},%Department and Organization
            addressline={}, 
            city={Seoul},
            postcode={04782}, 
            state={},
            country={South Korea}}

%% Abstract
\begin{abstract}
Accurate attenuation and scatter corrections are crucial in positron emission tomography (PET) imaging for accurate visual interpretation and quantitative analysis. Traditional methods relying on computed tomography (CT) or magnetic resonance imaging (MRI) have limitations in accuracy, radiation exposure, and applicability. Deep neural networks provide potential approaches to estimating attenuation and scatter-corrected (ASC) PET from non-attenuation and non-scatter-corrected (NASC) PET images based on VAE or CycleGAN. However, the limitations inherent to conventional GAN-based methods, such as unstable training and mode collapse, need further advancements. To address these limitations and achieve more accurate attenuation and scatter corrections, we propose a novel framework for generating high-quality ASC PET images from NASC PET images: Generation-Prior Diffusion Model (GPDM). Our GPDM framework is based on the Denoising Diffusion Probabilistic Model (DDPM), but instead of starting sampling from an entirely different image distribution, it begins from a distribution similar to the target images we aim to generate. This similar distribution is referred to as the Generation-Prior. By leveraging this Generation-Prior, the GPDM framework effectively reduces the number of sampling steps and generates more refined ASC PET images. Our experimental results demonstrate that GPDM outperforms existing methods in generating ASC PET images, achieving superior accuracy while significantly reducing sampling time. These findings highlight the potential of GPDM to address the limitations of conventional methods and establish a new standard for efficient and accurate attenuation and scatter correction in PET imaging.
\end{abstract}

%%Graphical abstract
\begin{graphicalabstract}
\includegraphics{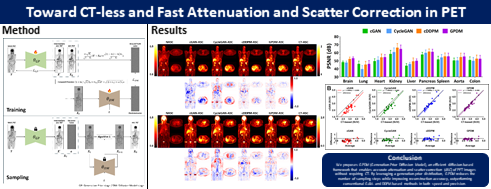}
\end{graphicalabstract}

%%Research highlights
\begin{highlights}
\item GPDM enables direct joint attenuation and scatter correction of PET.
\item Generation-prior accelerates diffusion sampling with fewer steps.
\item GPDM outperforms GAN and DDPM in NMSE, PSNR, and MAE metrics.
\item GPDM shows robust performance across scanners and TOF resolutions.
\end{highlights}

%% Keywords

\end{frontmatter}

%% Add \usepackage{lineno} before \begin{document} and uncomment 
%% following line to enable line numbers
%% \linenumbers

%% main text
%%

%% Use \section commands to start a section
\section{Introduction}
\label{sec:introduction}
To ensure accurate visual interpretation and quantitative analysis in positron emission tomography (PET) imaging, attenuation correction (AC) is essential. Conventionally, PET AC relies on structural imaging modalities, such as computed tomography (CT) or magnetic resonance imaging (MRI). CT-based AC methods are widely preferred for their simplicity and effectiveness, achieved through direct mapping of CT Hounsfield units to 511 keV linear attenuation coefficients~\cite{1}.

Conventional AC methods for PET imaging based on CT or MRI have several inherent limitations. In PET/CT, misregistration between CT and PET due to motion or respiratory effects is a significant source of inaccuracy in PET images~\cite{2}. Additionally, metal artifacts in CT images can introduce substantial errors in PET reconstructions~\cite{3}. There are also concerns that radiation exposure from CT scans may increase long-term health risks for pediatric patients~\cite{4}. Meanwhile, MRI-based AC encounters inherent challenges as there is no direct way to convert MRI signal intensity into 511 keV attenuation coefficients~\cite{5,6,7}. Moreover, the accuracy of clinically adopted MRI-based PET AC methods, such as MRI segmentation- and standard atlas-based approaches~\cite{8}, has been demonstrated only in normal adult brains to be within acceptable quantitative limits~\cite{9,10}. In addition, the emergence of CT-less brain-dedicated PET scanners demands PET AC methods that do not rely on anatomical images~\cite{11}.

To mitigate the limitations of conventional PET AC methods, many recent studies have increasingly adopted deep learning (DL) techniques. One of these DL-based approaches focuses on estimating pseudo-CT images or PET attenuation maps from non-attenuation-corrected (non-AC) PET or MRI images~\cite{12,13}. For instance, a convolutional neural network was utilized to synthesize pseudo-CT images from MRI, which demonstrated improved PET quantification accuracy~\cite{12}. A cycle-consistent generative adversarial network (CycleGAN) that generates pseudo-CT from non-AC PET has also yielded notable results in whole body studies ~\cite{13}. Another research direction involves directly estimating AC PET from non-AC PET images using DL models~\cite{14,15}. A U-Net-based architecture was developed to predict AC PET from non-AC PET, showing strong agreement with standard methods~\cite{14}. Additionally, a hybrid adversarial network that combines adversarial training with cycle consistency has been proposed to enhance the quality of AC PET images derived from non-AC PET~\cite{15}.

\begin{figure*}[t]
    \centering
    \includegraphics[width=\textwidth]{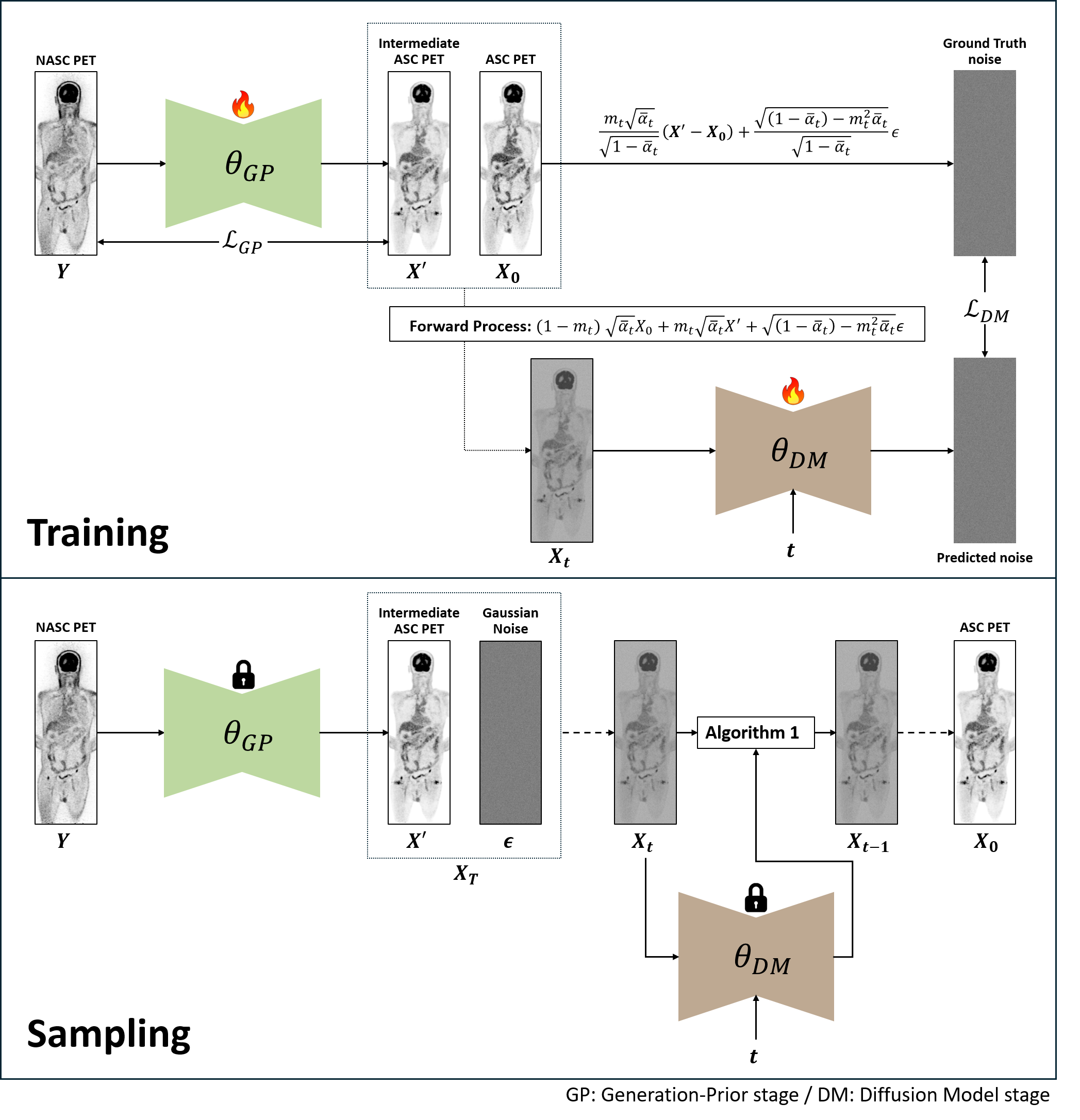} % 그림 파일 이름
    \caption{Overall Framework for Training and Sampling in GPDM}
    \label{fig:example}
\end{figure*}

In addition, recent studies have proposed simultaneously addressing AC and scatter correction (SC) using DL techniques. This integrated approach reduces computational costs and mitigates errors associated with treating these processes independently by leveraging the interdependent nature of attenuation and scatter in PET imaging ~\cite{16,17,18,shiri2023decentralized,shiri2020deep}. For instance, Yang et al. developed a DL model that takes non-AC and non-SC PET data as input to perform AC and SC simultaneously, achieving comparable or superior quantitative accuracy to traditional methods~\cite{16}. Similarly, Sun et al. utilized a GAN to jointly perform AC and SC for multi-tracer total-body PET, enhancing both the quality and quantitative accuracy of PET images~\cite{17}.

Recently, Denoising Diffusion Probabilistic Models (DDPMs)~\cite{19,20} have demonstrated remarkable performance, outperforming traditional U-Net and GAN-based models in natural image generation tasks. This advanced approach has also shown promising potential in PET image processing~\cite{21,22}. While previous DL approaches for PET AC or joint attenuation and scatter correction (ASC) has predominantly relied on GAN-based methods, these models pose limitations, such as training instability and challenges in accurately capturing complex image distributions~\cite{23,24}. Motivated by these advancements and existing gaps, we aim to develop a DDPM-based model specifically designed for joint ASC in PET imaging by leveraging the inherent strengths of DDPMs in generating high-fidelity and diverse outputs.

In this study, we propose a fast and efficient conditional probability-based DDPM to directly generate ASC PET images from non-attenuation- and non-scatter-corrected PET (NASC PET) images. While DDPMs share structural similarities with Variational Autoencoders (VAEs), their performance advantage lies in learning the transformation trajectory across multiple steps rather than relying on a single-step mapping, as seen in cGANs. However, conventional DDPMs must initiate sampling from a distribution entirely different from the target image, requiring approximately 1000 iterative sampling steps to generate an image. This method results in substantial computational overhead. To address these challenges, we proposed the Generation-Prior Diffusion Model (GPDM). GPDM introduces a framework that provides the Diffusion Model with a Generation-Prior, an image distribution similar to the target image, and utilizes it as the basis for sampling to generate the target image. To implement this approach, an Intermediate ASC PET image is first generated from NASC PET in a single step using a generative model. Next, a DDPM is applied to refine the Intermediate ASC PET, producing the ASC PET image. During the training process, the networks corresponding to the  Generation-Prior stage and  the  Diffusion Model stage are optimized alternately. Once training is complete, the network in the Generation-Prior stage is first used to produce the prior representation $x'$, which is then combined with Gaussian noise and progressively refined by the diffusion network to generate the final ASC PET image. This approach aims to reduce sampling time while enhancing performance. The proposed method was validated using whole-body \textsuperscript{18}F-FDG PET data through both organ-level and voxel-wise comparisons with CT-based ASC (CT-ASC) PET images. Furthermore, comparative evaluation with existing generative models, including cGAN and CycleGAN, demonstrated the superior accuracy of the proposed GPDM framework.

\section{Materials and Methods}
Our framework, depicted in Figure 1, aims to generate highly accurate ASC PET from NASC PET. It consists of two main components: a model for generating Intermediate ASC PET from NASC PET and a DDPM for transforming Intermediate ASC PET into ASC PET. First, we provide a brief introduction to DDPM, followed by a detailed explanation of GPDM for generating ASC PET from Intermediate ASC PET.

\subsection{Denoising Diffusion Probabilistic Models}
A T-step DDPM consists of a diffusion process and its reverse process. The diffusion process transforms the initial data \(x_0\) into \(x_T\)  over \(T\) steps using a fixed Markov chain and Gaussian models~\cite{19,20}:
\begin{equation}
q(x_1, \dots, x_T \mid x_0) = \prod_{t=1}^T q(x_t \mid x_{t-1}),
\end{equation}
where \(
q(x_t \mid x_{t-1}) = \mathcal{N}\left(x_t; \sqrt{1 - \beta_t}x_{t-1}, \beta_t I\right)
\) is Gaussian model with variance \(\beta_t\). One of important properties of the forward process \(q\) is that it can be sampled at any step, using \(x_0\) as the initial data:
\begin{equation}
q(x_t \mid x_0) = \mathcal{N}\left(x_t; \sqrt{\bar{\alpha}}x_0, (1 - \bar{\alpha})I\right) \end{equation}
\[x_t = \sqrt{\bar{\alpha}}x_0 + \sqrt{1 - \bar{\alpha}}\epsilon, \quad 
\epsilon \sim \mathcal{N}(0, I)\] where \(
\alpha_t = 1 - \beta_t \quad \text{and} \quad \bar{\alpha}_t = \prod_{s=1}^t \alpha_s
\). In other words, during the diffusion process, Gaussian noise is added to the previous image at each of the \(T\) steps, following a pre-defined schedule of beta values. This gradually converts the image at the initial time point, \(t=0\), into an image at time \(T = 1000\), which follows an isotropic Gaussian distribution.

During the reverse process of Eq.(1), a neural network is trained to predict the mean \(\mu_\theta(x_t, t)\) and variance \(\Sigma_\theta(x_t, t)\) of the image at time \(t-1(x_{t-1})\) given the image at time \(t (x_t)\) and time step \(t\) at each time:
\begin{equation}
p_\theta(x_{t-1} \mid x_t) = \mathcal{N}\left(x_{t-1}; \mu_\theta(x_t, t), \Sigma_\theta(x_t, t)\right).
\tag{3}
\end{equation}

The training objective for model (3), derived from the evidence lower bound (ELBO) concept, is formulated as follows~\cite{19}:
\begin{align}
\mathbb{E} \left[ \log p_\theta(x_0) \right] 
&\leq \mathbb{E}_q \left[ - \log \frac{p_\theta(x_{0:T})}{q(x_{1:T} \mid x_0)} \right] \tag{4} \\
&= \mathbb{E}_q \bigg[ D_{\mathrm{KL}}(q(x_T \mid x_0) \| p(x_T)) \notag \\
&\quad + \sum_{t \geq 1} D_{\mathrm{KL}}(q(x_{t-1} \mid x_t, x_0) \| p_\theta(x_{t-1} \mid x_t))  - \log p_\theta(x_0 \mid x_1) \bigg]. \notag
\end{align}

Ho \textit{et al.} [19] proposed that predicting the cumulative noise \(\epsilon_t\) added to the image \(x_t\) at time \(t\) is the best technique to parameterize this model. Thus, the network is trained using a simplified loss function instead of Eq.(4):
\begin{align}
L_t &= \mathbb{E}_{x_t, \epsilon_t}\left[\|\epsilon_t - \epsilon_\theta(x_t, t)\|^2\right] \notag \\
&= \mathbb{E}_{x_t, \epsilon_t}\left[\|\epsilon_t - \epsilon_\theta(\sqrt{\bar{\alpha}}x_0 + \sqrt{1 - \bar{\alpha}}\epsilon_t, t)\|^2\right].
\tag{5}
\end{align}

Consequently, we can derive the subsequent parameterization of the predicted mean \(\mu_\theta(x_t, t)\), which predicts the mean of the image at \(t-1 (x_{t-1})\) by subtracting the predicted Gaussian noise \(\epsilon_\theta(x_t, t)\) in the image at \(t\):
\begin{equation}
\mu_\theta(x_t, t) = \frac{1}{\sqrt{\alpha_t}} 
\left( x_t - \frac{\beta_t}{\sqrt{1 - \bar{\alpha}_t}} \epsilon_\theta(x_t, t) \right). 
\tag{6}
\end{equation}

In other words, DDPM uses a step-by-step diffusion process to transform the target image we want to generate into an isotropic Gaussian distribution. In this reverse process, they progressively restore the target image by estimating and eliminating the noise at each stage, which means denoising.

\subsection{Generation-Prior Diffusion Model fo Direct PET Attenuation and Scatter Corrrection}
The proposed GPDM consists of two main stages: the Generation-Prior stage and the Diffusion Model stage. In the Generation-Prior stage, a U-Net-based network is employed to learn the mapping between NASC PET and ASC PET, producing an intermediate output referred to as the Intermediate ASC PET. This intermediate output serves as the input for the subsequent Diffusion Model stage. The network is trained using a Mean Squared Error (MSE) loss function, minimizing the pixel-wise difference between NASC PET and ASC PET to effectively capture their relationship.

\begin{algorithm} 
\caption{Sampling for ASC PET}
\setstretch{1.3}
\textbf{Require:} $T, y, \{ \alpha_t \}_{t=1}^T, \{ m_t \}_{t=1}^T$

$x' \gets GP_\theta(y)$ (Generation Prior)

$z \sim \mathcal{N}(0, I)$

$x_T = \sqrt{\alpha_T} x' + \sqrt{1 - 2\alpha_T} z$

\textbf{for} $t = T$ \textbf{to} $1$

\hspace{0.5cm}    $\hat{\varepsilon}_t \gets DM_\theta(\alpha_t, x', t)$
    
 \hspace{0.5cm}   $\zeta_t \gets \frac{(1 - m_t)(1 - \bar{\alpha}_{t-1} - m_{t-1}^2 \bar{\alpha}_{t-1})}{(1 - m_{t-1})(1 - \bar{\alpha}_t - m_{t}^2 \bar{\alpha_t})} \sqrt{\alpha_t}$
    
  \hspace{0.5cm}  $\hat{x}'_{t-1} \gets \left(m_t \zeta_t + \frac{(1 - m_{t-1})}{\sqrt{\alpha_t}}\right)x_t$

  \hspace{0.5cm}  $\hat{x}'_{t-1} \gets  \hat{x}'_{t-1} + \left(m_{t-1} \sqrt{\bar{\alpha}_{t-1}} + m_t \sqrt{\bar{\alpha_t}}\zeta_t\right)x'$
    
  \hspace{0.5cm}  $x_{t-1} \gets \hat{x}'_{t-1} - \sqrt{1 - \bar{\alpha}_t} \left(\frac{1 - m_{t-1}}{\sqrt{\alpha_t}} - (1 - m_t) \zeta_t \right)\hat{\varepsilon}_t$
    
\textbf{end for}

\textbf{return} $x_0$

\end{algorithm}

In the diffusion and reverse processes of the diffusion model, traditional DDPM assumes isotropic Gaussian noise. However, in GPDM, the noise added at each step is modeled as a combination of isotropic Gaussian noise and systematic noise, rather than pure isotropic Gaussian noise, to incorporate the Generation-Prior. The systematic noise is defined as the difference between ASC PET and Intermediate ASC PET, addressing the residual correction required. If the Generation-Prior is not used, this systematic noise instead reflects the relationship between NASC PET and ASC PET. A neural network is trained to predict both noise components at each step of the reverse process, enabling the recovery of ASC PET by progressively removing these noise components from Intermediate ASC PET.

Within the framework of the diffusion processes in the Diffusion Model stage, an interpolation parameter \(m_t\) is utilized to merge the ASC PET image \(x_0\) with the Intermediate ASC PET image \(x'\), which is represented as the combination of the ASC PET image \(x_0\) and the systematic noise \(n\) (\(x' = x_0 + n\)). We can newly define the diffusion process as follows:
\begin{align}
q(x_t \mid x_0, x') &= \mathcal{N}\bigg( x_t; 
(1 - m_t)\sqrt{\bar{\alpha}_t}x_0 \notag \\
&\quad + m_t\sqrt{\bar{\alpha}_t}x', \left((1 - \bar{\alpha}_t) - m_t^2\bar{\alpha}_t\right)I
\bigg). 
\tag{7}
\end{align}
where \((1 - \bar{\alpha}_t) - m_t^2\bar{\alpha}_t\) is the variance. In (7), we replace the Gaussian mean in (2) with a linear combination between the ASC PET \(x_0\) and Intermediate ASC PET \(x'\) using the combination ratio \(m_t\). During the diffusion process from the ASC PET to intermediate ASC PET, over the \(T\) steps, \(m_t\) starts at 0 and gradually approaches 1, which is a predefined schedule.
At the final time point \(T\), unlike the original DDPM, the \(x_T\) image is a combination of isotropic Gaussian noise and Intermediate ASC PET \(x'\).

The reverse process starts from \(x_T\), which is Intermediate ASC PET \(x'\) with variance \(1 - 2\bar{\alpha}_T\) and \(m_T = 1\):
\begin{equation}
p_\theta(x_T \mid x') = \mathcal{N}(x_T; \sqrt{\bar{\alpha}_T}x', (1 - 2\bar{\alpha}_T)I).
\tag{8}
\end{equation}
Similar to Eq. (3), the neural network is trained to predict \(x_{t-1}\) based on \(x_t\) and \(x'\):
\begin{equation}
p_\theta(x_{t-1} \mid x_t, x') = \mathcal{N}(x_{t-1}; \mu_\theta(x_t, x', t), \Sigma_\theta(x_t, x', t)).
\tag{9}
\end{equation}
The model is trained to estimate \(\mu_\theta(x_t, x', t)\) through the modification of the existing method with the following ELBO formulation:
\begin{align}
\mathbb{E}_q \Big[ &D_{\mathrm{KL}}(q(x_T \mid x_0, x') \| p(x_T \mid x')) + \notag \\
&\sum_{t \geq 1} D_{\mathrm{KL}}(q(x_{t-1} \mid x_t, x_0, x') \| p_\theta(x_{t-1} \mid x_t, x')) - \log p_\theta(x_0 \mid x_1, x') \Big].
\tag{10}
\end{align}

In a manner similar to Ho \textit{et al.}'s approach, we considered predicting the cumulative mixture noise, comprising Gaussian noise and systematic noise added to the image \(x_t\) at the current time point \(t\), as the most effective technique for parameterizing this model. Thus, the network is trained using a simplified loss function instead of (10):
\begin{align}
L_t &= \mathbb{E}_{x_t, x_0, x', \epsilon_t} \Bigg[ 
\Bigg\| \Bigg( \frac{m_t \sqrt{\bar{\alpha}_t}}{\sqrt{1 - \bar{\alpha}_t}} (x' - x_0) + \frac{\sqrt{(1 - \bar{\alpha}_t) - m_t^2 \bar{\alpha}_t}}{\sqrt{1 - \bar{\alpha}_t}} \epsilon \Bigg)  \notag \\
&- \epsilon_\theta(x_t, x', t) \Bigg\|_2^2 \Bigg]. 
\tag{11}
\end{align}

In summary, the network of the Generation-Prior stage is trained to minimize the difference between the ASC PET and the generated image. Additionally, the difference between the ASC PET and the Intermediate ASC PET is combined with Gaussian noise and fed into the Diffusion Model stage. The network predicts the combined noise added at each step and is trained to minimize the difference between this prediction and the actual added noise.

The GPDM framework is trained using the following Value Function. During the training process, the networks corresponding to the Generation-Prior stage and the Diffusion Model stage are optimized alternately, with each training iteration sequentially updating one network at a time:
\begin{align}
\text{DM}^{(k+1)} = \arg\min_{\text{DM}} \, \mathcal{L}_{\text{DM}}, \
\text{GP}^{(k+1)} = \arg\min_{\text{GP}} \, \mathcal{L}_{\text{GP}}  \tag{12}
\end{align}
\begin{align}
\mathcal{L}_{\text{DM}} = \mathbb{E} \left[ \left\| \epsilon_t - \text{DM}(x_t, x', t) \right\|_2^2 \right], \quad x' = \text{GP}^{(k)}(y)
\tag{13} 
\end{align}
\begin{align}
\mathcal{L}_{\text{GP}} = \mathbb{E} \left[ \left\| x_0 - \text{GP}(y) \right\|_2^2 \right]
\tag{14}
\end{align} where GP refers to the network utilized in the Generation-Prior stage while DM denotes the network employed in the Diffusion Model stage. In Eq. (12), \(y\) represents NASC PET, \(x_0\) denotes ASC PET which is based on CT, and \(x'\) refers to the predicted image generated by the Generation-Prior stage network as the Generation-Prior for Diffusion Model stage.

In the image sampling process, the network of the Generation-Prior stage is first used to produce the Generation-Prior (\(x'\)). Sampling in the Diffusion Model stage then begins with a combination of the Generation-Prior and Gaussian noise. Through an iterative process, noise is gradually removed using the trained network, ultimately generating the ASC PET image. Specifically, the function \(\epsilon_\theta(x_t, x', t)\) is employed to compute the combined noise, which is subtracted from the joint distribution of \(x_t\) and \(x'\) to obtain the refined data \(x_{t-1}\). This process is repeated over \(T\) steps to produce the ASC PET, \(x_0\).

Original DDPM methods start from Gaussian noise and require a large number of sampling steps. In contrast, our approach leverages the similarity between the distributions of the ASC PET and the Generation-Prior, significantly improving sampling efficiency. Consequently, the sampling process is reduced from the typical 1000 steps to 200 steps. The impact of the number of sampling steps on performance will be discussed in detail in subsequent sections.

\begin{figure}[!t]
\centerline{\includegraphics[width=\columnwidth]{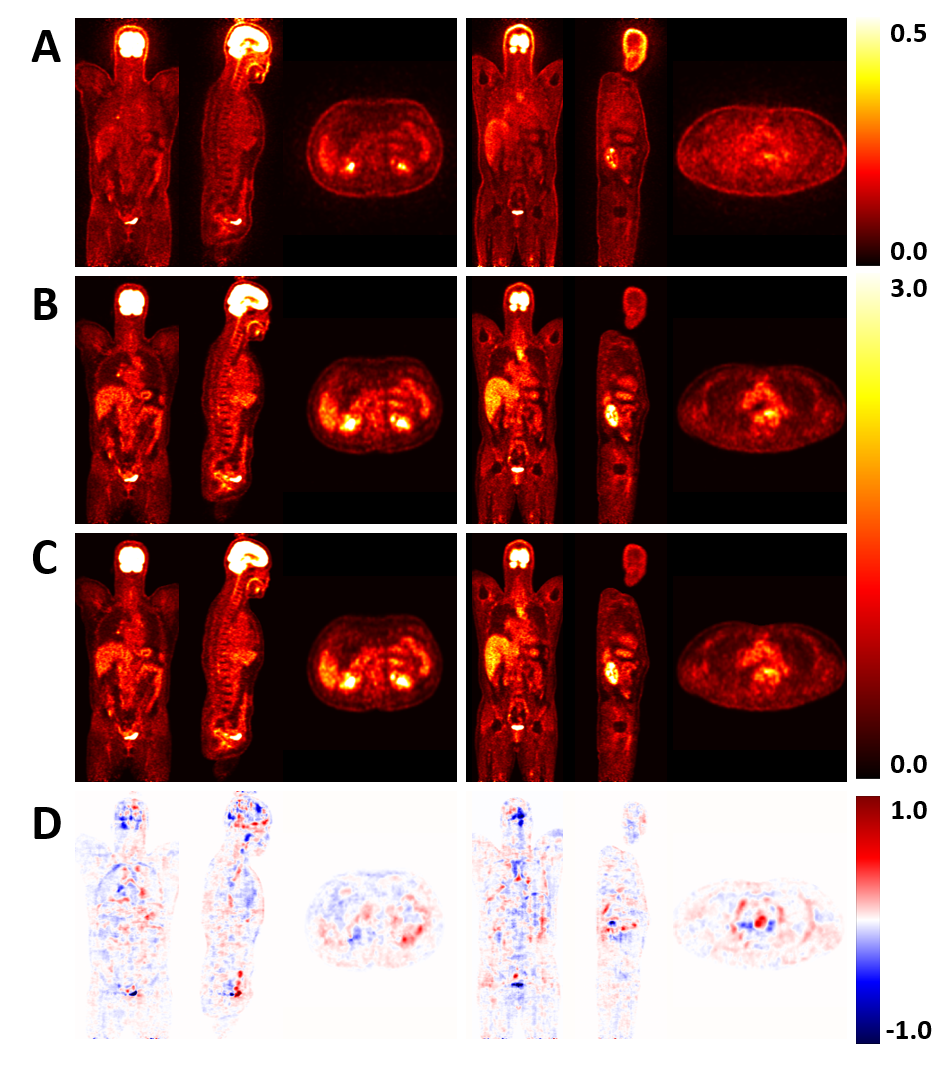}}
\caption{Attenuation and scatter corrected PET images of two representative patients using (A) non-ASC and (B) CT-based ASC, (C) GPDM, and (C) their difference. The unit of the color bar is SUV.}
\label{fig2}
\end{figure}

\subsection{Subjects and Image Acquisition}
In this study, a retrospective analysis was conducted on the whole-body \textsuperscript{18}F-FDG PET/CT scan data of 85 oncologic patients, which were acquired using a Siemens Biograph mCT 40 scanner. Of these patient data, 40 were used for training the neural networks, 5 for validation, and the remaining 40 for testing the accuracy of ASC. After the intravenous injection of \textsuperscript{18}F-FDG at a dose of 5.18 MBq/kg, PET/CT imaging was performed 60 minutes later. A 6- to 8-bed-position emission scan covered the upper bodies of patients from the head to the upper thigh, with each bed position lasting one minute.

Both NASC and ASC PET images were generated by reconstructing static images using a 3D order-subset expectation maximization algorithm with time-of-flight information of the e7 toolkit. In this toolkit, attenuation correction is performed using CT-derived attenuation maps, and scatter correction is implemented through a single scatter simulation (SSS) approach. The matrix size of PET images was \(200 \times 200 \times 109\) (\(4.07 \times 4.07 \times 2.03\) mm\textsuperscript{3} voxel size) for each bed position. Gaussian post-filter with a 4mm full-width at half-maximum was applied. The CT images were reconstructed into a \(512 \times 512 \times 100\) matrix with a voxel size of \(1.52 \times 1.52 \times 2.03\) mm\textsuperscript{3}. These were then converted into a \(\mu\)-map for 511-keV photons with a resolution of \(200 \times 200 \times 109\) and a voxel size of \(4.07 \times 4.07 \times 2.03\) mm\textsuperscript{3}~\cite{29}.

\subsection{Implementation Details}
The proposed method was implemented on the PyTorch open-source platform and executed using the NVIDIA GeForce GTX 3090 graphics processing unit (GPU). The generator employed in this study was based on the U-Net architecture~\cite{26}, renowned for its efficacy in medical image applications~\cite{25,27,28,29,30,31}. For training the network, the Adam optimizer was employed with a learning rate of 0.0001, and the maximum number of training epochs was set to 200. 

To mitigate the computational demands for DDPM, all comparative experiments evaluating the performance of the proposed method were conducted using a 2.5-dimensional training approach. In order to address the issues with in-plane discontinuities and incorporate more information from nearby slices, we utilized the 2.5D U-net which is a hybrid model that combines features from the 2D U-net and 3D U-net. When a patient data has dimensions of 200 × 200 × 109 for each bed, slicing it along the axial direction for 2D training would result in a network input of size 200 × 200 × 1. However, for 2.5D training, the network input is expanded to a total size of 200 × 200 × 7 by including three slices above and below the desired axial slice as neighboring slices. Consequently, the output is provided as a result of size 200 × 200 × 7, incorporating the desired axial slice from AC PET as well as adjacent slices above and below. The input data were standardized by subtracting the average value and adjusting the scale to achieve unit variance. All the networks compared in this study underwent identical training using a 2.5D model comprising seven slices. This uniform training approach ensured consistency across the models evaluated in the study. 

\begin{figure}[h]
\centerline{\includegraphics[width=\columnwidth]{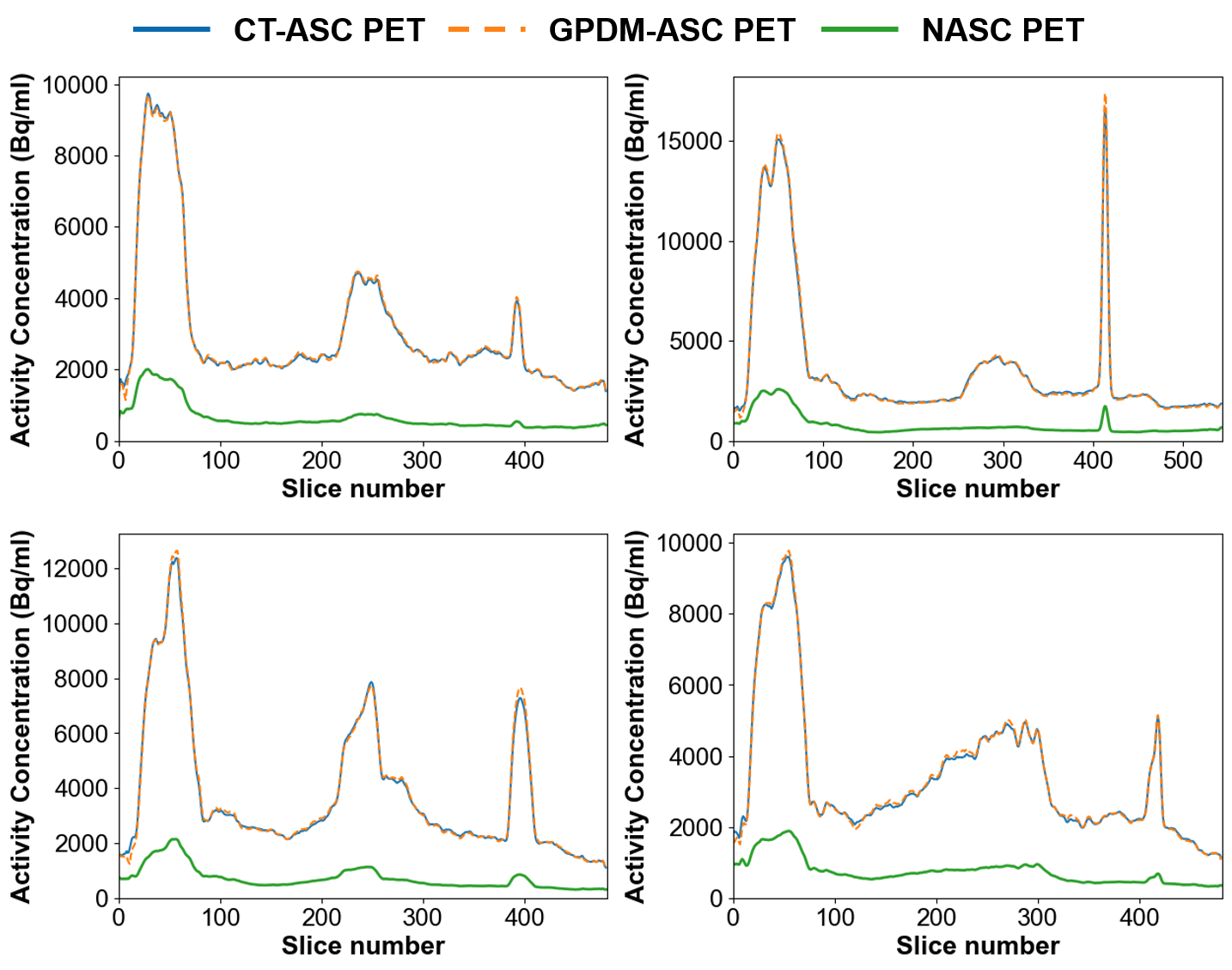}}
\caption{Cranial–caudal profiles of four representative patients, obtained by plotting the summed activity concentration across each axial (z-direction) slice for NASC PET, CT-ASC PET, and GPDM-ASC PET images.}
\label{fig3}
\end{figure}

\section{Results}
\subsection{Experimental Results}
To evaluate the reliability of the proposed method, ASC PET images generated by the proposed approach were compared with those generated by existing methods. For this purpose, metrics such as mean absolute error (MAE), normalized mean squared error (NMSE), and peak signal-to-noise ratio (PSNR) were employed:

\begin{align}
\text{MAE} &= \frac{\sum_{i \in V} |I_{\mathrm{AC}}(i) - I_{\mathrm{DL-AC}}(i)|}{\sum_{i \in V} I_{\mathrm{AC}}(i)} \notag\\
\text{NMSE} &= \frac{\sum_{i \in V} \big(I_{\mathrm{AC}}(i) - I_{\mathrm{DL-AC}}(i)\big)^2}{\sum_{i \in V} I_{\mathrm{AC}}(i)^2} \notag \\ 
\text{PSNR} &= 10 \log_{10} \left( \frac{N \cdot \max_{i \in V} I_{\mathrm{AC}}(i)}{\sum_{i \in V} \big(I_{\mathrm{AC}}(i) - I_{\mathrm{DL-AC}}(i)\big)^2} \right) \notag
\end{align}
where \(i\) is the voxel inside the contoured organs or whole body volume \(V\), and \(N\) is the total number of voxels inside the volumes. We also compared the $\mathrm{SUV}_{\mathrm{mean}}$ of bone lesions (12 lesions from 2 patients) and soft-tissue lesions (31 lesions from 11 patients) obtained by manual drawing of regions of interest on the CT-based ASC PET images.

These metrics were calculated between CT-based ASC PET (CT-ASC PET) and those generated by each method on the validation dataset. The calculations were performed for the entire body volume and for contoured organs, including the brain, lungs, heart, left and right kidneys, liver, pancreas, spleen, aorta, and colon. Organ segmentation was conducted using TotalSegmentator~\cite{32}.

The ASC PET images obtained through the proposed method (GPDM) closely resembled CT-ASC PET images. Stacking generated 2.5D slices into a 3D volume did not introduce observable overlap artifacts. Figure 2 compares the GPDM-ASC and CT-ASC SUV images of two representative patients. The proposed method exhibited low activity error in the lung regions except for their boundary, which is remarkably different from previous direct ASC PET estimation methods with high lung activity errors. The difference in the lung boundary would be attributed to CT-ASC error due to the respiratory lung motion. The cranial-caudal profiles of four other representative patients shown in Figure 3 also demonstrate the remarkable similarity of GPDM-ASC PET to the ground truth CT-ASC. This observation was further validated through entire test set of 40 patients: MAE = 6.02\% ± 1.52\%, NMSE = 0.83\% ± 0.40\%, and PSNR = 52.3 ± 5.2 dB.

\begin{table}[h]
    \centering
    \caption{Quantitative comparison of ASC PET images generated through cGAN, CycleGAN, cDDPM, and GPDM relative to CT-ASC PET.}
    \label{tab:comparison}
    \renewcommand{\arraystretch}{1.5} % 행 간격 조정
    \setlength{\tabcolsep}{4pt} % 열 간격 조정
    \resizebox{\columnwidth}{!}{ % 표를 열 폭에 맞춤
    \begin{tabular}{lcccc}
        \hline
        \textbf{Method} & \textbf{cGAN } & \textbf{CycleGAN} & \textbf{cDDPM} & \textbf{GPDM} \\ 
        \hline
        MAE (\%)       & 11.56 (2.08)      & 11.27 (1.63)        & 6.85 (2.38)       & \textbf{6.15 (1.37)} \\
        NMSE (\%)    & 3.17 (1.02)      & 2.41 (0.63)        & 1.19 (0.97)       & \textbf{0.84 (0.40)} \\
        PSNR (dB)       & 46.23 (4.62)       & 47.37 (5.13)        & 51.15 (5.04)       & \textbf{52.12 (5.06)} \\
        $p$-value (vs. GPDM)    & $p < 0.001$
 &$p < 0.001$
 & $p < 0.01$
  \\
         Sampling Time (s)            & 22.58  & 17.63 & 2158.23 & 449.79 \\
        \hline
    \end{tabular}
    }
\end{table}

\begin{figure*}[t]
    \centering
    \includegraphics[width=\textwidth]{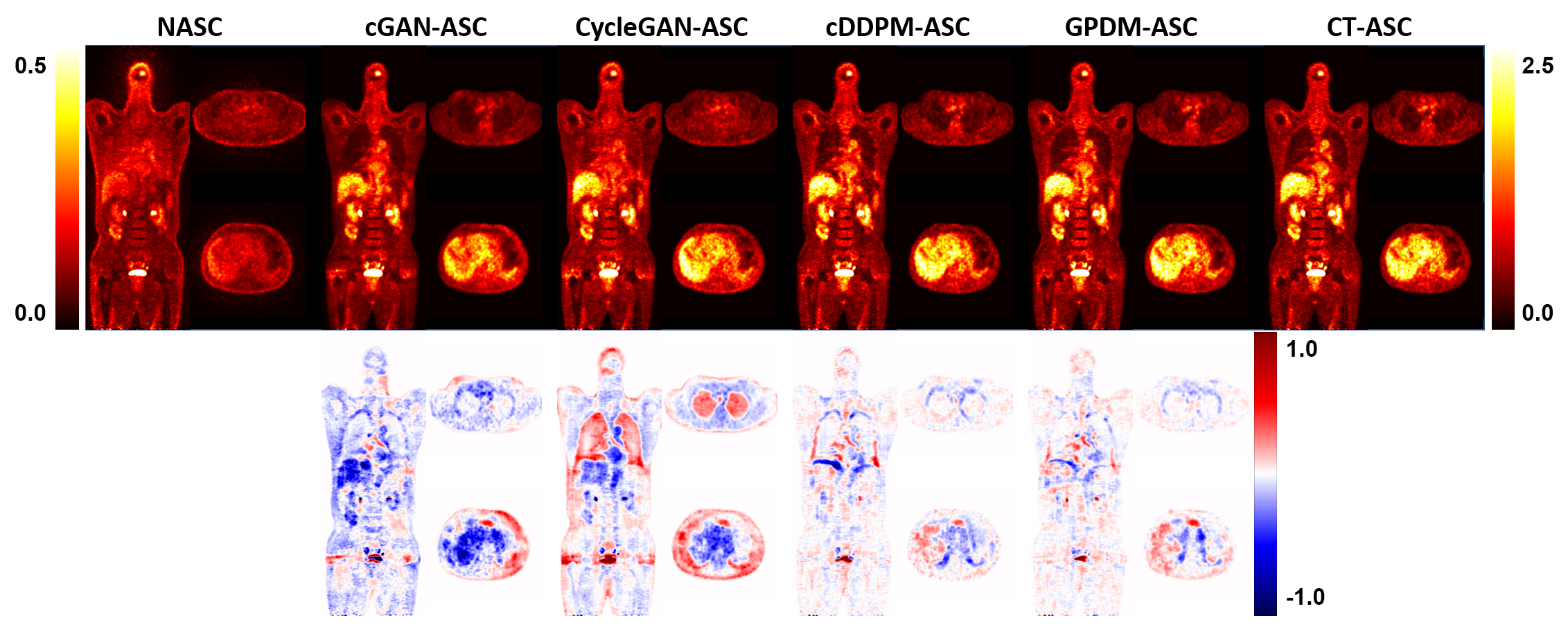} % 그림 파일 이름
    \caption{Comparison of ASC PET images generated through cGAN, Cycle GAN, cDDPM and GPDM for one patient. The error maps relative to CT-ASC PET are shown in the second row. The unit of the color bar is SUV.}
    \label{fig:example}
\end{figure*}

\begin{figure*}[t]
    \centering
    \includegraphics[width=\textwidth]{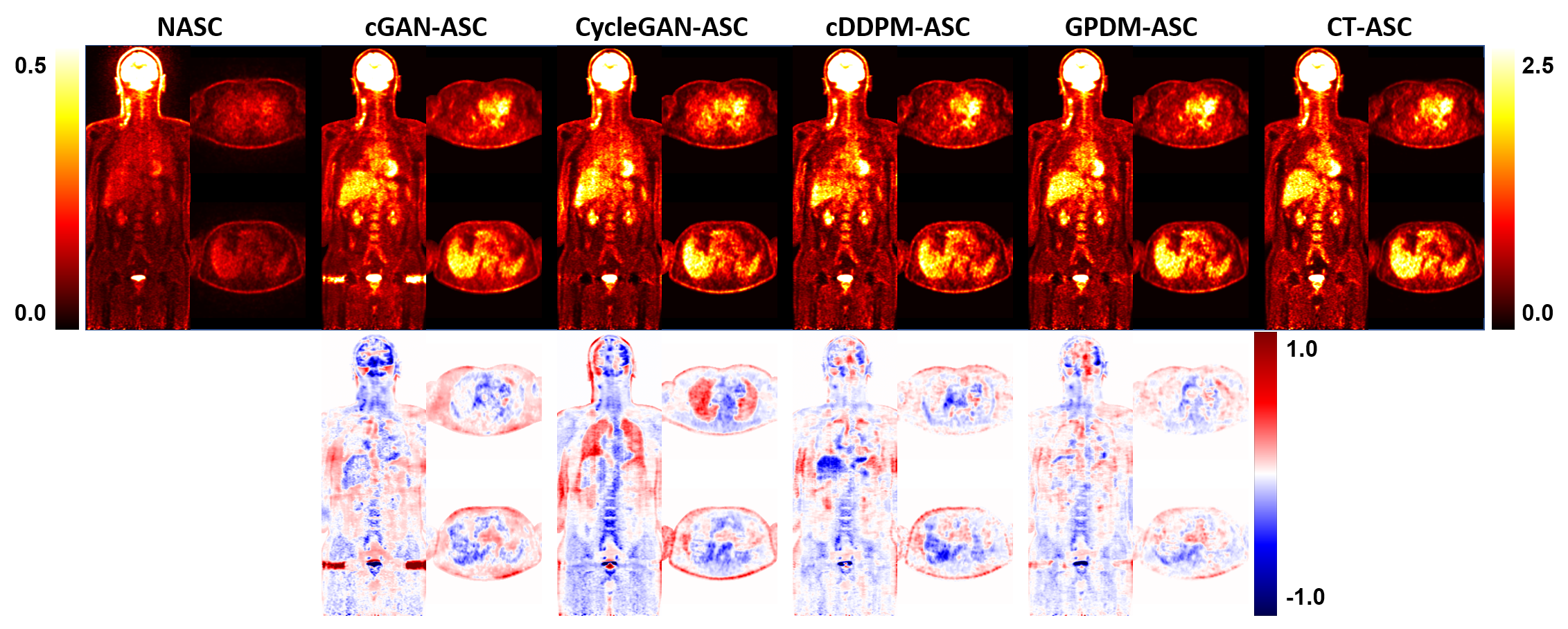} % 그림 파일 이름
    \caption{. Comparison of ASC PET images generated through cGAN, Cycle GAN, cDDPM, and GPDM for a different patient. The error maps relative to CT-ASC PET are shown in the second row. The unit of the color bar is SUV.}
    \label{fig:example}
\end{figure*}

The proposed method outperformed the conventional generative models. Table 1 presents the quantitative metrics for the whole body, comparing GPDM with cGAN, CycleGAN, and cDDPM which refers to generation of ASC PET directly from NASC PET y in the Diffusion Model stage without utilizing the Generation-Prior \(x\)’. The proposed GPDM demonstrated substantial improvements over conventional generative models in all key metrics. Specifically, in terms of MAE, GPDM outperformed cGAN and CycleGAN by approximately 47\% and 45\%, respectively, and showed a 10\% improvement over cDDPM. Similarly, GPDM achieved approximately 74\% and 65\% lower NMSE values compared to cGAN and CycleGAN, respectively, and was 30\% superior to cDDPM. In PSNR, GPDM surpassed cGAN and CycleGAN by approximately 13\% and 10\%, respectively, and outperformed cDDPM by 2\%. Although the sampling time of GPDM was longer than that of one-step generative models like cGAN and CycleGAN, it was significantly faster than cDDPM, which requires 1000 sampling steps. These results highlight the overall superiority of GPDM in both performance and efficiency. Figures 4 and 5 compare the coronal and axial ASC PET images corrected using different approaches, along with their error maps relative to the CT-ASC PET. Figure 4 particularly highlights the superior performance of proposed DDPM methods in the lung and upper abdomen compared to cGAN and CycleGAN. This superiority was further confirmed through the voxel-wise correlation analysis results summarized in Table 2. GPDM exhibited the highest R² values among all the models.

\begin{table}[h]
    \centering
    \caption{Voxel-wise correlation of ASC PET images generated through cGAN, CycleGAN, cDDPM, and GPDM relative to CT-ASC PET.}
    \label{tab:correlation}
    \renewcommand{\arraystretch}{1.5} % 행 간격 조정
    \setlength{\tabcolsep}{6pt} % 열 간격 조정
    \resizebox{\columnwidth}{!}{ % 표를 열 폭에 맞춤
    \begin{tabular}{lccc}
        \hline
        \textbf{Method} & \textbf{Slope} & \textbf{Intercept} & \textbf{R\textsuperscript{2}} \\ 
        \hline
        cGAN            & 0.901 (0.021)   & 0.039 (0.034)       & 0.961 (0.015) \\
        CycleGAN        & 0.968 (0.021)   & 0.047 (0.017)       & 0.966 (0.012) \\
        cDDPM           & 0.986 (0.012)   & 0.010 (0.018)       & 0.983 (0.012) \\
        GPDM            & \textbf{0.992 (0.009)} & \textbf{0.007 (0.011)} & \textbf{0.988 (0.006)} \\
        \hline
    \end{tabular}
    }
\end{table}

Figure 6 presents a comparison of the quantitative metrics of the proposed GPDM against cGAN, CycleGAN, and cDDPM at the organ level, highlighting the superior performance of the proposed method across nearly all organs. Notably, the NMSE values for the brain, heart, kidney, liver, spleen, pancreas, and aorta consistently remained below 1.3\%. Furthermore, while the PSNR between conventional generative models and CT ASC fell below 45 dB in some organs, the proposed method consistently exceeded this threshold, demonstrating its robustness and reliability.

As shown in Figure 7, a similar trend was observed in the regional $\mathrm{SUV}_{\mathrm{mean}}$ quantification for both bone and soft tissue lesions. Among the four methods (cGAN, CycleGAN, cDDPM, and GPDM), the proposed GPDM model showed the highest consistency with CT-based attenuation correction. For bone lesions, GPDM produced a regression line of $y = 1.007x - 0.032$, and for soft tissue lesions, $y = 1.049x - 0.268$. These regression lines are very close to the identity line ($y = x$), indicating that GPDM closely matches the SUV values derived from CT-based correction. The Bland--Altman plots further support this result, showing that GPDM produced the smallest differences and the most stable patterns across both tissue types. While cDDPM also showed a strong correspondence with CT-based SUVs, its performance was slightly lower than that of GPDM. In contrast, cGAN and CycleGAN resulted in larger bias and more variation, especially in bone lesions, suggesting less reliable performance.

\subsection{Ablative Studies}
\subsubsection{Comparison of Different Models for Generation-Prior}
In our approach, we employed a simple supervised learning method using a U-Net to generate the Generation-Prior. In this section, we evaluate the performance of alternative generative models for this task. Specifically, we replaced the U-Net with cGAN for Generation-Prior, training it with the value function of GPDM as follows:
\begin{align}
&\min_{\mathrm{GP}} \max_D \min_{\mathrm{DM}} V(\mathrm{GP}, D, \mathrm{DM}) & \notag \\
&= \mathbb{E}\big[\log D(x_0, y)\big] + \mathbb{E}\big[\log\big(1 - D(\mathrm{GP}(y), y)\big)\big] & \tag{15} \\
&\quad + \mathbb{E}\Big[\big\|\epsilon_t - \mathrm{DM}(x_t, \mathrm{GP}(x), t)\big\|_2^2\Big]. & \notag
\end{align}

Figure 7 shows the results of employing U-Net, cGAN, and the case where no generative model was used (cDDPM) for comparison. The quantitative evaluation for whole-body performance is summarized in Table 3. Similar to U-Net~\cite{26}, cGAN~\cite{33} demonstrated improved performance compared to the absence of a generative model. As shown in Table 3, both the U-Net-based GP model (UNet-GP) and the cGAN-based GP model (pix2pix-GP) improved the overall image quality compared to the case without a generative prior. While their overall performance is comparable, the U-Net-based model achieved slightly better results across all metrics: MAE (6.02\% vs. 6.18\%), NMSE (0.83\% vs. 0.87\%), and PSNR (52.31~dB vs. 52.09~dB).

\begin{figure}[!h]
\centerline{\includegraphics[width=\columnwidth]{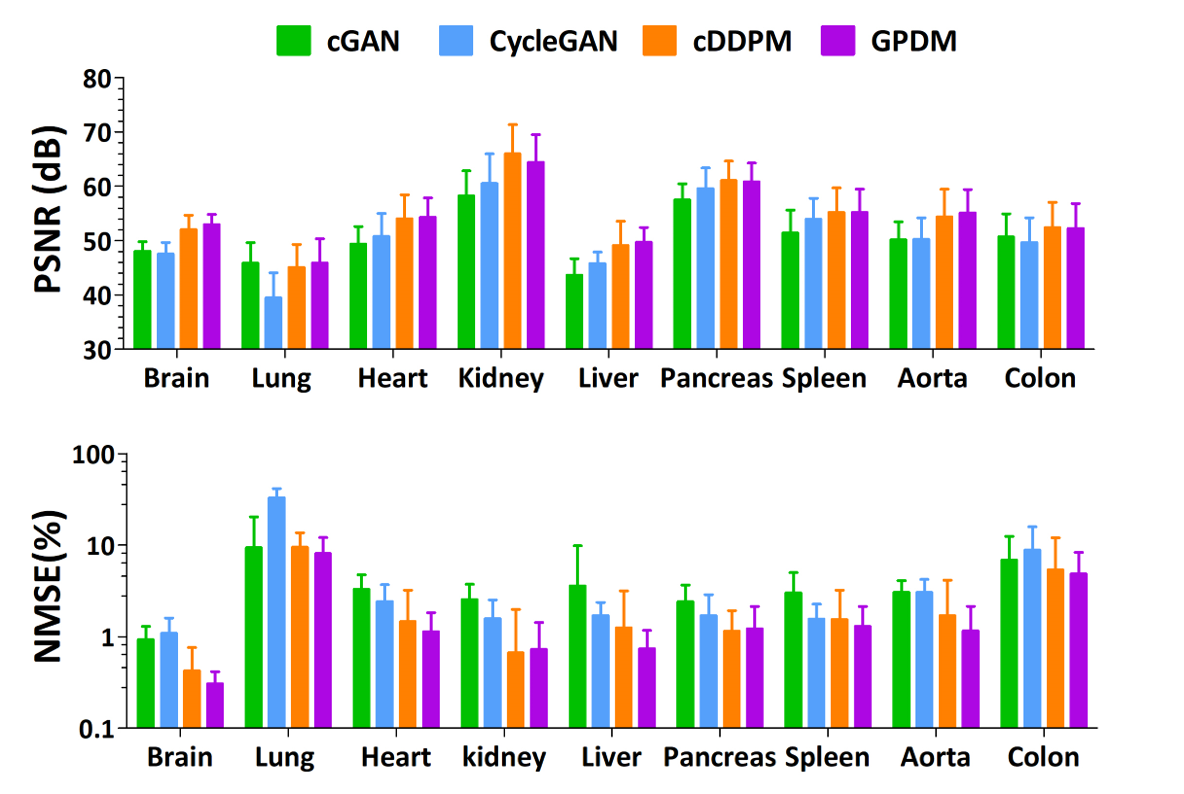}}
\caption{PSNR and NMSE of ASC PET images using generative models relative to CT-ASC PET.}
\label{fig6}
\end{figure}

\begin{figure}[!h]
\centerline{\includegraphics[width=\columnwidth]{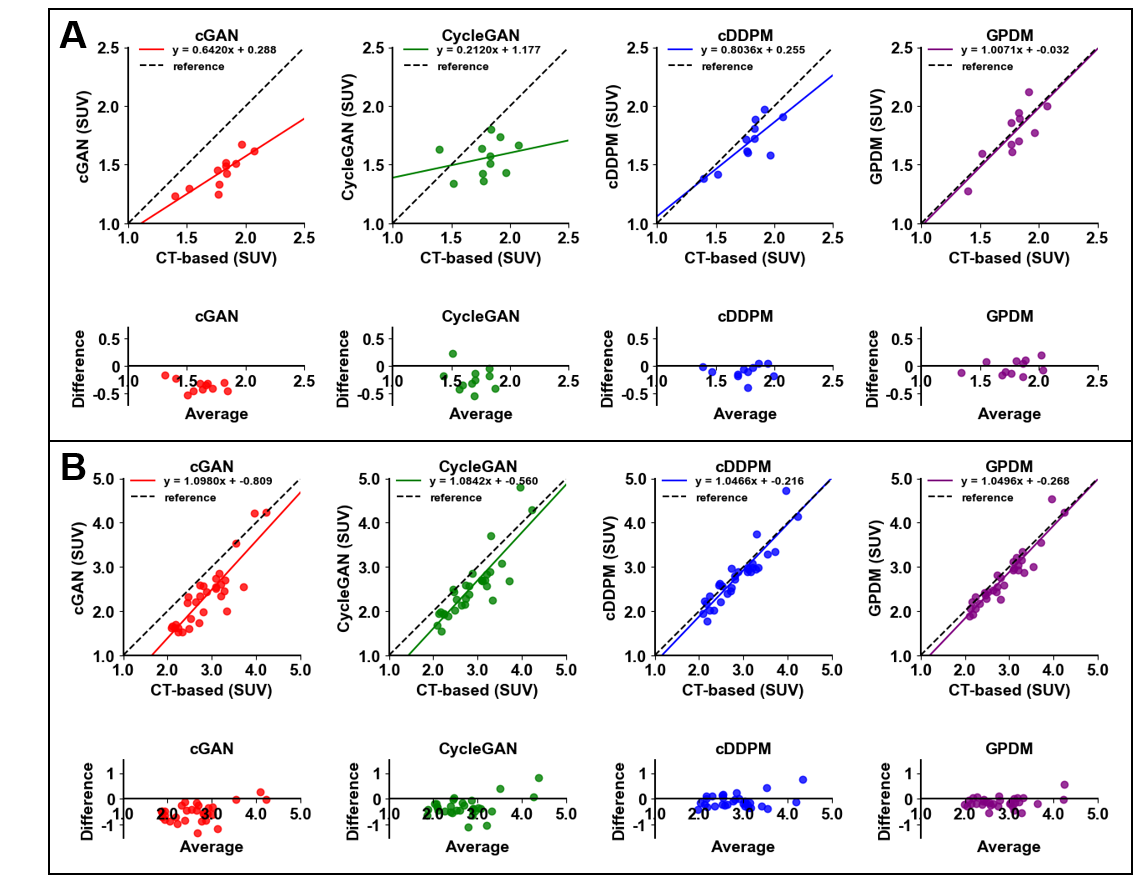}}
\caption{Comparison of SUV mean measured in (A) bone lesions and (B) soft tissue lesions}
\label{fig7}
\end{figure}

\subsubsection{Impact of Sampling Step on performance}
In our primary method, the number of steps in the diffusion model was set to 200. To assess the influence of this parameter, we additionally evaluated configurations with 500 and 100 steps. The quantitative results of this analysis are summarized in Figure 8. While variations in the number of steps led to minor differences in performance, the overall impact was minimal.

\subsubsection{Stability Across Randomized Training Splits}
To evaluate the robustness of our framework against variations in training data, we conducted an ablation study using four different random training subsets, each consisting of 40 patients selected from the 80 patients available for training. A fixed validation set of 5 patients was used across all experiments. For each training subset, we independently trained the proposed model and assessed its performance on the same validation set. As shown in Table~\ref{tab:split_comparison_transposed}, the results were highly consistent across all splits. To determine whether the differences among the four models were statistically significant, we performed a Friedman test on the NMSE values. The resulting p-value was 0.78, indicating no significant difference ($p > 0.05$). These results suggest that our framework is robust to random variations in the training data composition.

\subsubsection{Robustness to TOF Resolution Differences}
The mCT dataset was acquired with a TOF resolution of 540 ps, whereas the Vision PET scanner offers a finer TOF resolution of approximately 210 ps. Because attenuation correction is highly sensitive to TOF information, we examined whether the proposed method remains effective under different TOF resolutions. To this end, the model was trained and evaluated on the Vision PET data, which provides higher TOF timing precision. As summarized in Table 5, the model achieved an NMSE of 0.31, PSNR of 62.24 dB, and MAE of 1.15, demonstrating that improved TOF resolution results in more accurate attenuation and scatter correction.

\section{Discussion}
In our study, we have shown the feasibility of the DDPM for generating ASC PET directly from NASC PET without the use of structural information. This method produces accurate attenuation distribution estimates even in the absence of structural information and demonstrates a high level of concordance when compared to CT-ASC PET. These results substantiate the high reliability of our model. The proposed method serves as an effective means to substitute or supplement ASC methods based on CT and MR imaging, eliminating the need for the acquisition of structural information for ASC. Additionally, this approach has the potential to mitigate quantitative biases arising from truncation artifacts in CT or MRI, spatial discrepancies between CT/MRI and PET imaging, and characteristic artifacts such as the ‘banana’ artifact commonly observed in the absence of proper attenuation correction.~\cite{34}.

\begin{table}[!t]
    \centering
    \caption{Quantitative results of generated ASC PET images using different models for the Generation-Prior (GP).}
    \label{tab:gp-results}
    \renewcommand{\arraystretch}{1.5} % 행 간격 조정
    \setlength{\tabcolsep}{6pt} % 열 간격 조정
    \resizebox{\columnwidth}{!}{ % 표를 단일 열 폭에 맞춤
    \begin{tabular}{lccc}
        \hline
        \textbf{Method} & \textbf{MAE (\%)} & \textbf{NMSE (\%)} & \textbf{PSNR (dB)} \\ 
        \hline
        w/o GP          & 6.73 (2.50)       & 1.18 (0.98)        & 51.32 (5.14) \\
        w/ UNet-GP      & \textbf{6.02 (1.52)} & \textbf{0.83 (0.40)} & \textbf{52.31 (5.16)} \\
        w/ pix2pix-GP   & 6.18 (1.59)       & 0.87 (0.45)        & 52.09 (5.18) \\
        \hline
    \end{tabular}
    }
\end{table}

\begin{table}[!t]
\centering
\caption{Performance comparison of the proposed framework across four different random training splits.}
\begin{tabular}{lcccc}
\hline
\textbf{Metric} & \textbf{Set 1} & \textbf{Set 2} & \textbf{Set 3} & \textbf{Set 4} \\
\hline
NMSE (\%)       & 0.846 & 0.808 & 0.839 & 0.808 \\
PSNR (dB)       & 49.75 & 49.93 & 49.76 & 49.87 \\
MAE (\%)         & 6.66 & 6.39 & 6.37 & 6.41 \\
Slope           & 1.0049 & 1.0017 & 1.0054 & 1.0049 \\
Intercept       & -0.00353 & -0.00194 & -0.00369 & -0.00243 \\
R$^2$           & 0.989 & 0.989 & 0.989 & 0.989 \\
\hline
\end{tabular}
\label{tab:split_comparison_transposed}
\end{table}

Conventional DDPM generate images by progressively removing Gaussian noise, but this formulation is not well suited for ASC PET. To better capture the relationship between NASC PET and ASC PET, we reformulated the noise process to incorporate both Gaussian noise and task-specific systematic noise. This design enables the diffusion model to progressively transform NASC PET into ASC PET in a task-oriented manner. Nevertheless, directly generating ASC PET from NASC PET still requires a large number of sampling steps because of the substantial distributional gap between the two domains. To overcome this inefficiency, we introduce a Generation-Prior, which serves as an intermediate representation closer to the target ASC PET distribution. By combining this prior with Gaussian and systematic noise at the start of the diffusion process, our model achieves more accurate reconstruction with significantly fewer sampling steps. Taken together, GPDM offers both improved accuracy and computational efficiency, establishing a practical and scalable solution that addresses the inherent limitations of CT/MR-dependent and pseudo-CT–based ASC methods.

The approach of generating pseudo-CT images using deep learning and subsequently using them for attenuation and scatter correction followed by image reconstruction has been extensively explored in previous studies. The strength of the pseudo-CT–based method is its compatibility with established reconstruction algorithms and its ability to model attenuation and scatter physics accurately. However, this approach requires an extra reconstruction step, which increases computational cost, can propagate pseudo-CT errors through the reconstruction process, and may be more sensitive to domain shifts. In contrast, directly predicting ASC-PET images from NASC-PET removes the need for intermediate pseudo-CT generation and subsequent reconstruction, enabling significantly faster processing and simpler clinical workflows. Moreover, it allows the network to learn accurate scanner-specific or systematic correction patterns that would otherwise need to be estimated by computationally expensive physics-based methods such as Monte Carlo simulation. The trade-offs are that the direct approach does not inherently guarantee data consistency with the measured events and has lower interpretability compared to the pseudo-CT approach. In this work, we selected the direct approach to achieve a fast and efficient correction process suitable for high-throughput clinical use, and GPDM compensates for its primary limitation by producing outputs that are highly consistent with standard ASC-PET images.

The conventional GAN approach trains simultaneously the generator and discriminator networks to perform the mapping from source images to target images ~\cite{23,33}. However, noise existing between input sources and output targets sometimes hinders the generator from learning meaningful image mappings. To address this issue, approaches like CycleGAN have been introduced. CycleGAN adopts a cyclical approach, adding inverse transformations to impose additional constraints on the generator~\cite{35}. This helps prevent mode collapse and assists the generator in finding unique and meaningful image mappings. Nevertheless, CycleGAN requires bidirectional mappings for transformations between two domains, which can lead to an unnecessary network learning process which is converting from ASC PET to NASC PET.

\begin{figure}[!t]
\centerline{\includegraphics[width=\columnwidth]{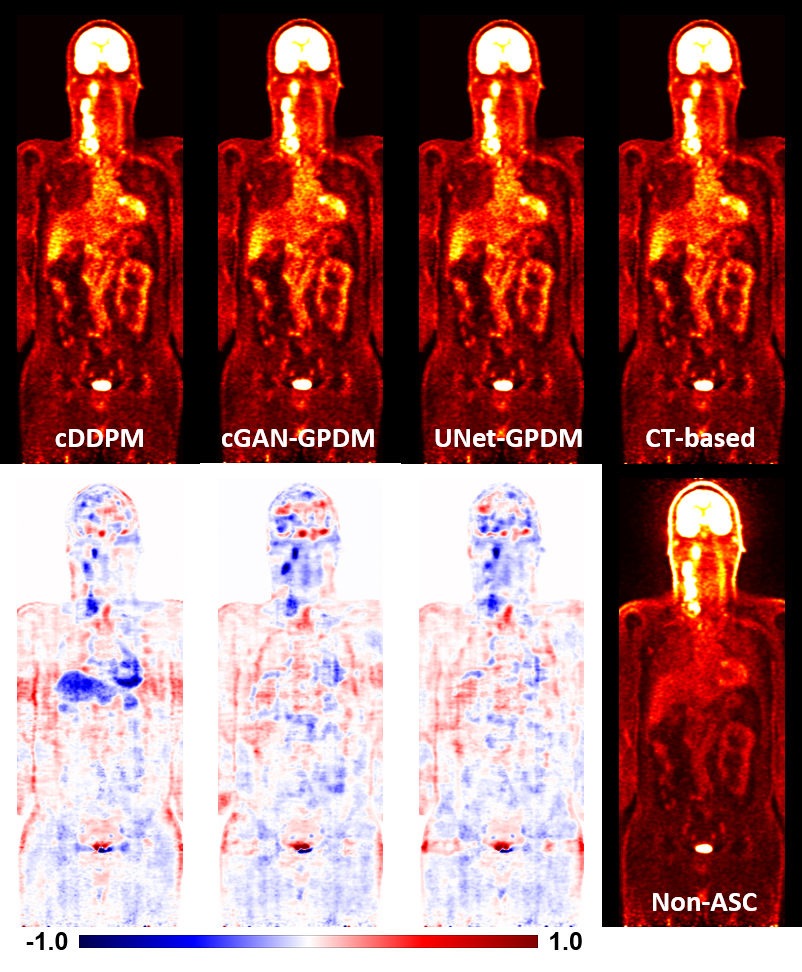}}
\caption{Visual comparison of generated ASC PET images using different models for the Generation-Prior.}
\label{fig7}
\end{figure}

\begin{table}[!t]
\centering
\caption{Quantitative evaluation of the proposed GPDM framework on datasets with different TOF resolutions.
The mCT dataset (540 ps) and Vision PET dataset (210 ps) show that the model achieves higher reconstruction accuracy with improved TOF timing resolution.}
\begin{tabular}{lcc}
\hline
Metric &  mCT &  Vision \\
\hline
NMSE (\%) & 0.84 & 0.31 \\
PSNR (dB) & 52.12 & 62.24 \\
MAE (\%) & 6.15 & 1.15 \\
\hline
\end{tabular}
\label{tab:vision_transposed}
\end{table}

Recently, DDPM has been introduced as a method that addresses the limitations of conventional generative models, demonstrating exceptional performance in image generation tasks. DDPM learns the transformation trajectory between the target image and Gaussian noise, generating images directly from Gaussian noise. Although this approach achieves high performance, it requires a large number of sampling steps due to its reliance on generating images from Gaussian noise.

To overcome these challenges, we propose GPDM. In conventional cDDPM, the sampling process starts with an image formed by combining NASC PET and Gaussian noise and progressively removes the combined noise to generate the ASC PET. However, a significant limitation remains here: the distribution of images formed by combining NASC PET and Gaussian noise differs substantially from that of ASC PET, necessitating a large number of sampling steps.

To address this limitation, the GPDM introduces a Generation-Prior into the Diffusion Model. Instead of directly generating ASC PET from NASC PET, the model generates ASC PET based on the Generation-Prior. The combined image of the Generation-Prior and Gaussian noise exhibits a distribution closer to that of ASC PET compared to the combination of NASC PET and Gaussian noise. This proximity enables the model to achieve higher performance with significantly fewer sampling steps. As a result, the required number of sampling steps was substantially reduced from 1000 to 200.

\begin{figure}[!t]
\centerline{\includegraphics[width=\columnwidth]{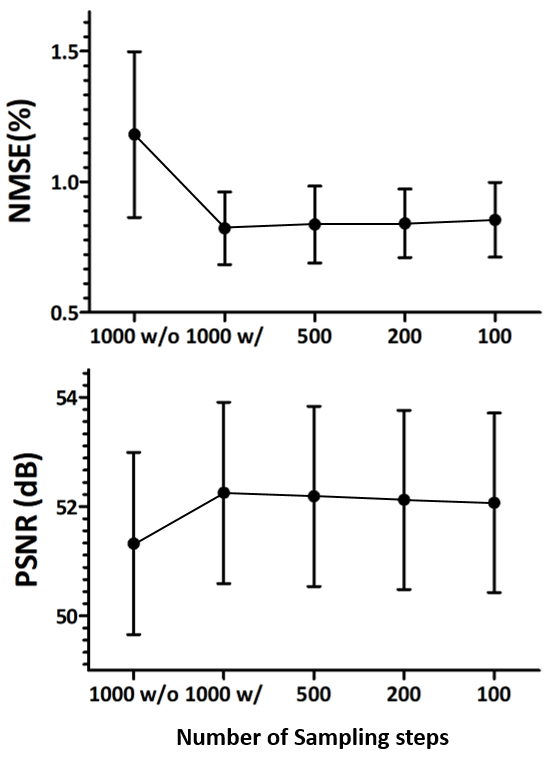}}
\caption{Quantitative analysis of the impact of the number of sampling steps on GPDM. The x-axis represents the number of sampling steps, with the 1000-step case indicating the absence of a Generation-Prior.}
\label{fig9}
\end{figure}

\begin{figure}[!h]
\centerline{\includegraphics[width=\columnwidth]{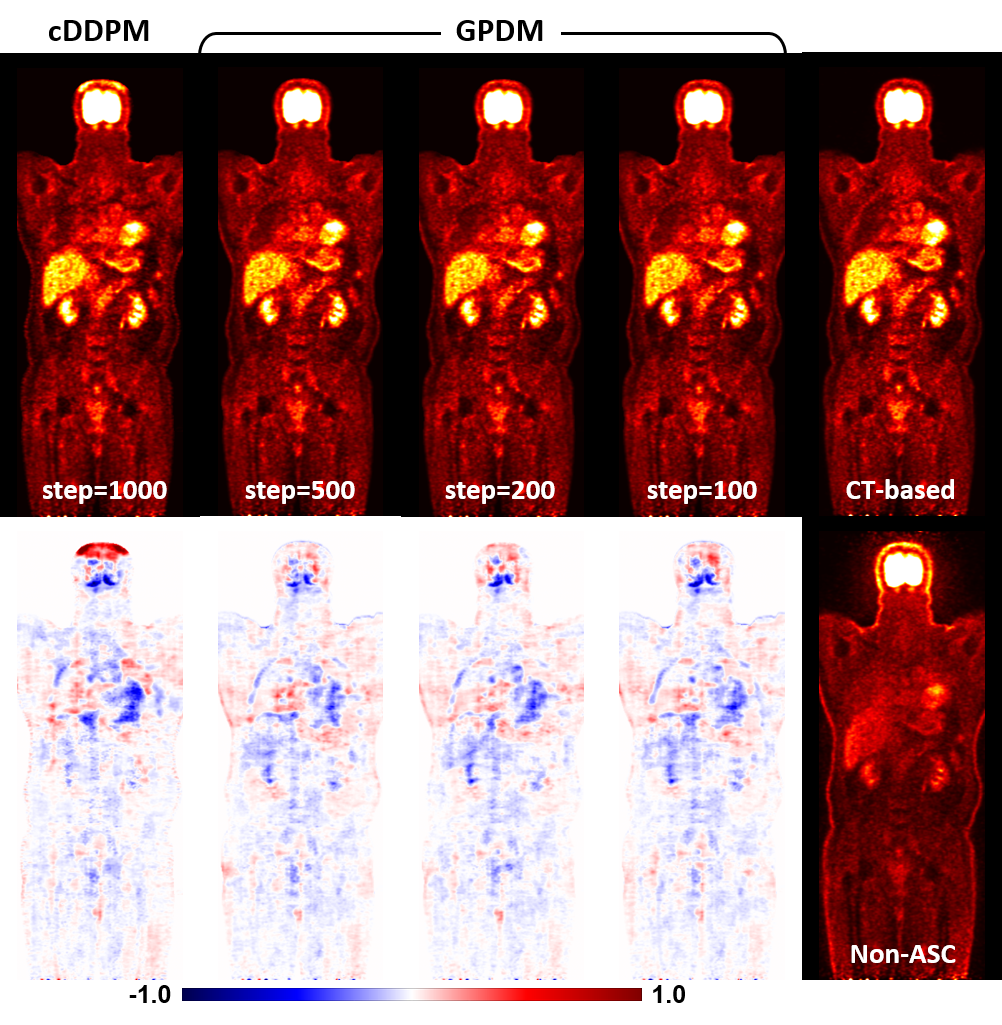}}
\caption{Visualization of generated ASC PET images using GPDM with varying sampling steps (500, 200, 100), alongside CT-based ASC PET and Non-ASC PET for comparison. The 1000-step case represents where a Generation-Prior is not used. The bottom row displays error maps showing differences relative to the CT-based ASC.}
\label{fig10}
\end{figure}

Although DDPM is related to VAE~\cite{36}, it generates higher-quality images by performing transformations over multiple steps. In addition, dividing transformations between two domains into multiple steps has been shown to improve performance in image transformation tasks~\cite{19,37}. In light of this, the concept of Generation-Prior was introduced. The quantitative assessment summarized in Table 1 demonstrates the superior performance of our proposed method over traditional generative models. The reduction in MAE and MSE, as well as the increase in PSNR, attests to the efficacy of our approach in learning the intricate relationship between NASC PET and ASC PET. The voxel-wise correlation analysis summarized in Table 2 and the organ-level analysis shown in Figure 6 also highlight that our proposed method outperforms conventional approaches. In addition, the regional $\mathrm{SUV}_{\mathrm{mean}}$ analysis for both bone and soft tissue lesions (Figure 7) supports these results. Among all compared methods, GPDM yielded the most consistent results with CT-based attenuation correction, with regression lines closest to the identity line. The Bland-Altman analysis also showed the smallest bias and the narrowest limits of agreement, indicating that the proposed method can provide reliable and consistent SUV measurements across different tissue types.

We conducted additional experiments to investigate the impact of the method used for producing the Generation-Prior on overall performance, yielding the results summarized in Table 3. Specifically, we tested two approaches: a basic supervised learning (U-Net) and a generative adversarial method (cGAN). The results indicate minimal performance differences between the two approaches, suggesting that the choice of network for generating the Generation-Prior has a relatively minor effect on performance.

Additionally, we conducted an ablation study to investigate the impact of step size in the diffusion process on the performance of our method. As shown in Figure 8, the use of Generation-Prior resulted in superior performance even with fewer steps compared to the case without Generation-Prior. Furthermore, while the larger step sizes yielded higher performance when Generation-Prior was used, the performance increment achieved by increasing step sizes were not substantial. Figure 9 presents an example of a representative patient, clearly demonstrating that Generation-Prior leads to more accurate ASC PET, particularly in the lung and heart regions, compared to when Generation-Prior is not used.

The ablation study using different random subsets of the training data confirmed that our framework delivers highly consistent performance even with a limited amount of in-distribution data, indicating strong generalizability within the same scanner domain. This suggests that the model can reliably learn attenuation and scatter correction patterns without being overly sensitive to specific patient compositions in the training set. However, PET attenuation and scatter correction are known to be highly sensitive to scanner-dependent factors such as time-of-flight (TOF) resolution and acquisition geometry. To examine whether the proposed framework remains effective under such variations, we conducted an additional experiment focusing on the robustness to TOF resolution differences. The mCT dataset was acquired with a TOF resolution of 540 ps, whereas the Vision PET scanner offers a finer TOF resolution of approximately 210 ps. To evaluate the impact of this difference, the model was trained and tested on the Vision PET data, which provides higher TOF timing precision. As summarized in Table 5, the model achieved an NMSE of 0.31, PSNR of 62.24 dB, and MAE of 1.15, demonstrating that improved TOF resolution leads to more accurate attenuation and scatter correction. These findings confirm that the proposed GPDM framework maintains strong performance across datasets with distinct TOF characteristics and effectively benefits from higher TOF precision.

Recent work by Xie et al.~\cite{xie2024dose} also employs a U-Net-based network to generate a prior for a diffusion model in the context of PET imaging, which is conceptually similar to our approach. However, their study focuses on PET image denoising, whereas our method is specifically designed for attenuation and scatter correction (ASC), which presents different challenges. Furthermore, while they directly applied the original DDPM framework after generating the prior, our method incorporates a newly designed DDPM structure that models task-specific systematic noise, making it more suitable for ASC in PET. This difference helps our diffusion model better handle the physical characteristics and specific needs of attenuation and scatter correction, emphasizing the novelty and practical focus of our method.

Once GPDM has completed its training, it can be used to generate ASC PET using only NASC PET as input. This trained model offers advantages, such as reducing radiation exposure and saving time during MRI acquisition, as there is no need to perform additional CT or MRI scans to predict ASC PET. Moreover, the trained model is not limited to a specific scanner; it can be applied to different PET scanners, enabling the generation of ASC PET even on PET scanners without CT or MR capabilities. This allows standalone PET scanners provide ASC PET results similar to those of PET/CT or PET/MR. However, it is important to note that using a model trained on a different PET scanner may yield compromised performance, requiring further validation studies. 

While our proof-of-concept study used a limited dataset due to the heavy computational load required for training GPDM, we acknowledge the potential for further improvement in performance with a larger dataset. In this study, we adopted a 2.5D training strategy, processing slices with adjacent contextual information rather than full 3D volumes, to reduce computational demands and enable effective learning from limited data. However, as photon scattering is inherently a 3D phenomenon, a full 3D training setup could potentially capture spatial correlations more comprehensively and improve correction accuracy. Such an extension would require significantly greater computational resources and larger datasets, which we consider an important direction for future work. The overshoot observed in the cDDPM around the top of the skull (Figure 10) may be partly attributed to the limitations of the 2.5D approach, and we expect that full 3D training could help reduce such artifacts. In contrast, GPDM did not exhibit similar overshoot artifacts despite being trained under the same 2.5D conditions. As further supported by the quantitative comparison in Figure 6, GPDM demonstrated lower errors in the brain region compared to cDDPM.

\section{Conclusion}
This study proposes a DDPM-based model for more accurately generating ASC PET from NASC PET, which is GPDM ~\cite{38}. The method introduces an Intermediate ASC PET, generated from NASC PET, which is then provided as a Generation-Prior to the DDPM. By transforming between distributions of two highly correlated images rather than entirely different domains, the model achieves improved performance and reduces the number of sampling steps required. This novel DDPM approach have demonstrated superior performance in terms of quantification accuracy and reliability. Furthermore, it is applicable to PET data collected from both standalone PET scanners and multimodal imaging systems such as PET/CT and PET/MRI.

%% The Appendices part is started with the command \appendix;
%% appendix sections are then done as normal sections
% \appendix
% \section{Example Appendix Section}
% \label{app1}

% Appendix text.

% %% For citations use: 
% %%       \cite{<label>} ==> [1]

% %%
% Example citation, See \cite{lamport94}.

%% If you have bib database file and want bibtex to generate the
%% bibitems, please use
%%
%%  \bibliographystyle{elsarticle-num} 
%%  \bibliography{<your bibdatabase>}

%% else use the following coding to input the bibitems directly in the
%% TeX file.

%% Refer following link for more details about bibliography and citations.
%% https://en.wikibooks.org/wiki/LaTeX/Bibliography_Management

\bibliographystyle{style} 
\bibliography{references}

% \begin{thebibliography}{00}

% %% For numbered reference style
% %% \bibitem{label}
% %% Text of bibliographic item

% %%\bibitem{lamport94}
%   %%Leslie Lamport,
%   %%\textit{\LaTeX: a document preparation system},
%   %%Addison Wesley, Massachusetts,
%   %%2nd edition,
%   %%1994.

% \end{thebibliography}
\end{document}